\documentstyle[prl,aps,amssymb,epsf]{revtex}

\begin{document}

\draft

\preprint{UTPT-97-05}

\title{Marginally Trapped Surfaces in 
the Nonsymmetric Gravitational Theory}

\author{L.\ Demopoulos and J.\ L\'egar\'e}
\address{Department of Physics, University of Toronto,
Toronto, Ontario, Canada M5S 1A7} 

\date{\today}

\twocolumn[
\maketitle

\widetext

\begin{abstract}%
We consider a simple, physical
approach to the problem of marginally trapped surfaces in
the Nonsymmetric Gravitational Theory ({\sc ngt}).
We apply this approach to a particular spherically symmetric, Wyman sector 
gravitational field, consisting of a pulse in the antisymmetric field
variable.
We demonstrate that marginally trapped surfaces do exist for this choice
of initial data.
\end{abstract}

]

\narrowtext

\section{The Metrics of Spherically Symmetric, Wyman Sector NGT}
\label{sec:metrics of NGT}

In general, the causal structure of {\sc ngt} is not known.
However, for the special case of spherically symmetric, Wyman sector
data, the causal structure of {\sc ngt}, whether old 
(see~\cite{bib:Moffat Phys Rev D 19 3554 1979,%
bib:Moffat Found Phys 14 1217  1984,%
bib:Moffat Banff}),
new 
(see~\cite{bib:Moffat Phys Lett B 355 447 1995,%
bib:Moffat J Math Phys 36 3722 1995,%
bib:Legare and Moffat Gen Rel Grav 27 761 1995}),
or dynamically-constrained
(see~\cite{bib:Moffat dynamical constraints})
collapses back to its Unified Field Theory ({\sc uft}) form
(see~\cite{bib:Einstein Ann Math 46 578 1945,%
bib:Einstein and Straus Ann Math 47 731 1946,%
bib:Lichnerowicz J Rat Mech and Anal 3 487 1954,%
bib:Lichnerowicz 1955,%
bib:Maurer-Tison C R Acad Sc 242 1127 1956,%
bib:Hlavaty 1958,%
bib:Maurer-Tison Ann scient Ec Norm Sup 76 185 1959,%
bib:Tonnelat 1982}).
The causal structure of {\sc uft} was investigated by 
F.~Maurer-Tison 
(see~\cite{bib:Maurer-Tison Ann scient Ec Norm Sup 76 185 1959});
there it was found that the field equations possessed three
(most likely distinct) characteristic surfaces, with 
metric components given by
\begin{equation}
\label{eq:Maurer-Tison metrics}
h_{\mu\nu} = g_{(\mu\nu)} , 
\quad l^{\mu\nu} = g^{(\mu\nu)} , 
\quad\hbox{\rm and}\quad 
\chi^{\mu\nu} = \frac{2h}{g} h^{\mu\nu} - l^{\mu\nu} .
\end{equation}
We use greek letters for spacetime indices and reserve latin letters
for spatial indices.
The $g_{\mu\nu}$ are
the components of the fundamental tensor, and $h$ and $g$ are the determinants
of the matrices formed by $h_{\mu\nu}$ and $g_{\mu\nu}$, respectively.
The fact that there are three metrics implies an ambiguity in
the measurement of geometrical quantities.
We see no {\it a priori} reason to choose any one metric over another,
and hence we will consider all three.

In spherically symmetric, Wyman sector {\sc ngt}, the fundamental tensor
takes the form
$\bbox{g}^{-1} = \bbox{e}_\bot \otimes \bbox{e}_\bot
- \phi^{-4}\gamma^{ij} \bbox{e}_i \otimes \bbox{e}_j$,
where the surface fundamental tensor is
\[
|\gamma^{ij}| =
\left[
\begin{array}{ccc}
\gamma^{11} &   &   \\
  & \gamma^{22} & \gamma^{[23]}/\sin\theta \\
  & -\gamma^{[23]}/\sin\theta & \gamma^{22}/\sin^2\theta \\
\end{array}\right] 
\]
and $\phi$ is the conformal factor, solution to the Hamiltonian constraint
(see~\cite{bib:Clayton Demopoulos and Legare 1997} for a discussion on
the solvability of the 
Hamiltonian constraint in {\sc ngt}).
The basis vectors form a surface-adapted frame, with the
$\bbox{e}_i$ forming a basis in a spatial hypersurface $\Sigma$, and 
$\bbox{e}_\bot$ being normal to this hypersurface.
These basis vectors 
are related to the holonomic coordinate frame by
$\bbox{\partial}_t = N\bbox{e}_\bot + N^i\bbox{e}_i$
and $\bbox{\partial}_i = \bbox{e}_i$, where $N$ is the lapse function
and $\bbox{N} = N^i\bbox{e}_i$ is the shift vector.
We take $\gamma^{22} = r^{-2}\cos\psi$ and $\gamma^{[23]} = r^{-2}\sin\psi$,
where $r$ is the radial coordinate and $\psi$ is an arbitrary function
characterizing the strength of the antisymmetric variables {\it versus}
the symmetric variables;
in particular, $\gamma^{[23]} / \gamma^{22} = \tan\psi$.
For the details of the Hamiltonian formulation of {\sc ngt} and 
its initial-value problem, we refer the reader to
the literature
(see~\cite{bib:Clayton Demopoulos and Legare 1997} in addition 
to~\cite{bib:Clayton thesis,bib:Clayton Int J Mod Phys D,%
bib:Clayton CQG 13 2851 1996}).

In this parametrization, it can be verified 
that~(\ref{eq:Maurer-Tison metrics}) are
$|h_{\mu\nu}| =
\mbox{\rm diag}[1, -\phi^4/\gamma^{11}, -\phi^4r^2\cos\psi,
-\phi^4r^2\cos\psi\sin^2\theta]$,
$|l_{\mu\nu}| =
\mbox{\rm diag}[1, -\phi^4/\gamma^{11}, -\phi^4r^2\sec\psi,
-\phi^4r^2\sec\psi\sin^2\theta]$,
and
$|\chi_{\mu\nu}| =
\mbox{\rm diag}[[\cos^2\psi - \sin^2\psi]^{-1}, 
-\phi^4 [\gamma^{11}[\cos^2\psi - \sin^2\psi]]^{-1},
-\phi^4r^2\sec\psi,
-\phi^4r^2\sec\psi\sin^2\theta]$.
We note in passing that if $\psi = \pi/4$, 
then this last metric is degenerate, while in a region
$\pi/4 - \epsilon < \psi < \pi/4 + \epsilon$ it changes signature.
These pathologies occur
despite the fact that $\psi \sim \pi/4$
does not correspond to a particularly strong field region in {\sc ngt},
since $\gamma^{22} \sim \gamma^{[23]}$.

\section{Marginally Trapped Surfaces in NGT}

\begin{figure}
\centerline{\epsffile{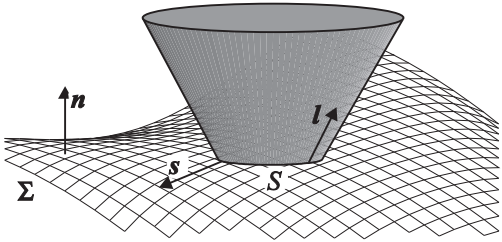}}
\medskip
\caption{A spatial slice $\Sigma$ of a spacetime is shown, with an
embedded surface $S$, represented here as a circle on $\Sigma$.
The timelike vector $\protect\bbox{n}$ 
is normal to $\Sigma$, while the spacelike vector $\protect\bbox{s}$
is normal to $S$, and is thus tangent to $\Sigma$.
The null vector $\protect\bbox{l}$ generates the evolution of $S$.
The surface $S$ is marginally trapped if the rate of change of its
area vanishes along $\protect\bbox{l}$: 
$\partial_{\protect\bbox{l}}[A] = 0$.}
\label{fig:marginally trapped surface figure}
\end{figure}

The standard definition of a marginally trapped surface 
in General Relativity is as follows.
Consider a spatial slice $\Sigma$ into which is embedded a 
2-surface $S$ (see Figure~\ref{fig:marginally trapped surface figure}, where
the 2-surface $S$ is represented as a circle).
Let $\bbox{n} = n^\mu \bbox{e}_\mu$ be the (timelike) normal 
vector field to $\Sigma$, 
and let
$\bbox{s} = s^\mu \bbox{e}_\mu$ be the (spacelike) normal vector field to $S$.
We therefore have $\bbox{n}\cdot\bbox{n} = 1$ 
and $\bbox{s}\cdot \bbox{s} = -1$, by proper
normalization.
It follows that $\bbox{l} = \bbox{n} + \bbox{s}$ is a null vector field:
$\bbox{l}\cdot \bbox{l} = 0$.
The 2-surface $S$ is a marginally trapped surface if the 
hypersurface generated by $\bbox{l}$ has vanishing expansion,
{\it i.e.}, if the trace of its extrinsic curvature vanishes.
This definition can be put into $3+1$ form by means of a simple argument
due to York (see~\cite{bib:York 1989}, p.~100).
Using the projection operator $\bbox{P}$ that projects a spacetime
quantity onto the surface generated by $\bbox{l}$,
\[
\bbox{P} = (\openone - \bbox{n}\otimes\bbox{n})
\cdot (\openone + \bbox{s}\otimes\bbox{s})
 = \openone - \bbox{n}\otimes\bbox{n} + \bbox{s}\otimes\bbox{s} 
\]
or
\[
P^\alpha_\beta = \delta^\alpha_\beta - n^\alpha n_\beta 
+ s^\alpha s_\beta ,
\]
then the extrinsic curvature of interest is 
$\bbox{\kappa} = - (\bbox{P}\otimes\bbox{P}^t)\cdot (\nabla[\bbox{l}])$
or $\kappa_\mu^\nu = - P_\mu^\alpha P_\beta^\nu\nabla_\alpha [l]^\beta$,
according to the usual definition.
The trace of $\bbox{\kappa}$ is the expansion of the surface generated by
$\bbox{l}$; making this trace vanish yields the {\em apparent
horizon equation}:
\begin{mathletters}
\label{eq:standard apparent horizon equation}
\begin{equation}
{}^{(3)}\nabla\cdot\bbox{s} - {\rm Tr}[\bbox{K}]
+ \bbox{K}(\bbox{s}, \bbox{s}) = 0
\end{equation}
or
\begin{equation}
{}^{(3)}\nabla_i[s]^i - K_i^i + K_{ij} s^i s^j = 0 .
\end{equation}
\end{mathletters}

It is difficult to generalize the above derivation to {\sc ngt}: it
relies on the definitions of both the extrinsic curvature and the
covariant derivative.
Presumably, some generalization can be given, however since both of
these concepts are ambiguously defined in {\sc ngt}, it is preferable
to approach the problem from a different perspective.

We again consider a spatial slice $\Sigma$ into which is embedded a
2-surface $S$.
Consider a null vector field $\bbox{l}$, where null is defined with respect
to one of the three metrics in~(\ref{eq:Maurer-Tison metrics}).
This vector field generates a one-parameter congruence 
off the surface $\Sigma$;
this parameter, which we denote by $\tau$, 
represents the time for ``observers'' 
flowing along the curves drawn out by $\bbox{l}$.
Let $A$ be the area of the 2-surface $S$.
Suppose we measure the area $A(\tau_0)$ at some $\tau = \tau_0$.
We then travel along the flow to some $\tau = \tau_0 + d\tau$, 
where $d\tau \ll \tau_0$, and
re-measure the area $A(\tau_0 + d\tau)$.
If $A(\tau_0 + d\tau) - A(\tau_0) = 0$, then the surface $S$ is
defined to be {\em marginally trapped}.
In other words, for infinitesimal $d\tau$,
\begin{equation}
\label{eq:geometric definition of apparent horizon}
\frac{A(\tau_0 + d\tau) - A(\tau_0)}{d\tau}
\rightarrow \left.\frac{dA(\tau)}{d\tau}\right|_{\tau = \tau_0}
= \partial_{\bbox{l}}[A] 
\buildrel {(!)} \over = 0 .
\end{equation}
Therefore, a marginally trapped surface has the property that its area
is unchanging along the flow generated by a null vector field $\bbox{l}$.
Note that (\ref{eq:geometric definition of apparent horizon}) contains
a partial derivative, since the area $A$ is, strictly speaking, a 
2-form:
$\partial_{\bbox{l}}[A] = \langle \bbox{d A},
\bbox{l}\rangle$, where $\bbox{A}$ is the area 2-form of $S$ and
$\bbox{d A}$ is its exterior derivative.

\section{A Simple Argument for the Existence of 
Marginally Trapped Surfaces in NGT}

For a spherically symmetric, Wyman sector 
gravitational field
it is a simple matter to demonstrate that the area of a surface of
constant $r$ is given by
$A(r) = 4\pi g_{22}$,
where $g_{22}$ is the $22$-component of one of the three metrics 
given above.
This gives us three possible values for the surface area
of a sphere of radius $r$ centred on the origin, two of which are equal:
\[
A_h(r) = 4\pi r^2\phi^4\cos\psi 
\]
and
\[
A_l(r) = A_\chi(r) = 4\pi r^2\phi^4\sec\psi .
\]
Here, $\phi(r)$ is the conformal factor, solution to the Hamiltonian 
constraint.
For a moment of time 
symmetry, (\ref{eq:geometric definition of apparent horizon})
simplifies considerably and we find that
a marginally trapped surface obtains if $\partial_r[A] = 0$.
We thus have three equations, two of which are identical, 
whose solution
gives the location of a marginally trapped surface:
\begin{mathletters}
\begin{equation}
\label{eq:marginally trapped surface h}
f_h(r) = 1 - \frac{M}{2r} + \partial_r[M] 
- \frac{r\tan\psi\partial_r[\psi]}{2}
\Bigl[1 + \frac{M}{2r}\Bigr] 
= 0 
\end{equation}
and
\begin{eqnarray}
\label{eq:marginally trapped surface l and chi}
f_l(r) = f_\chi(r) &=& 1 - \frac{M}{2r} + \partial_r[M] \nonumber \\
& & \quad\mbox{} + \frac{r\tan\psi\partial_r[\psi]}{2}
\Bigl[1 + \frac{M}{2r}\Bigr]
= 0 , 
\end{eqnarray}
\end{mathletters}%
where we have introduced the mass function $M(r)$, defined by
$\phi(r) = 1 + M(r) / 2r$.
For an asymptotically flat spatial slice, 
the mass function converges to a constant,
$M_{\rm ADM}$, the {\sc adm} mass of the system.
Setting 
$\psi = 0$ demonstrates that
these equations indeed reduce to their {\sc gr} counterpart:
$f_{\rm GR}(r) = 1 - M/2r + \partial_r[M] = 0$.

Consider an initial data set consisting of some arbitrary function 
$\gamma^{11}$ that falls off asymptotically 
(see~\cite{bib:Clayton thesis}, p.~116, for the precise requirement),
a function $\psi$, and a vanishing extrinsic curvature.
We assume that this initial data set forms a well-posed initial-value problem,
and hence that there exists a solution to the Hamiltonian
constraint, $\phi$.
The function $\psi$ is taken to be more or less localized about a 
point $r_0$ of the initial slice.
Thus, $\psi$ can be characterized by three parameters: an
overall amplitude factor $A$ that essentially serves to fix
the size of the {\sc adm} mass of the system, 
the position of its peak $r_0$, and
its width $\sigma$.
The exact form of $\psi$ is irrelevant: all that is required is that
$\sigma \ll r_0$.
In the regions $r \lesssim r_0 - \sigma$ and
$r \gtrsim r_0 +\sigma$, we assume that $\psi$ is negligible relative
to its peak value at $r = r_0$.
For such a data set, it is a relatively simple matter to demonstrate
that the mass function takes the form
(see~\cite{bib:Clayton Demopoulos and Legare 1997})
\[
M(r) \approx \left\{
\begin{array}{ll}
M_{\rm ADM} r / r_0 & \mbox{\rm for $r \lesssim r_0 - \sigma$, while} \\
M_{\rm ADM} & \mbox{\rm for $r \gtrsim r_0 + \sigma$.} \\
\end{array}
\right.
\]
The corrections are of the order $\sigma/r_0$, so that we require
that $\sigma/r_0 \ll 1$.
This approximation is most inaccurate in the region
$r_0 - \sigma < r < r_0 + \sigma$; however, since we assume that
$\sigma/r_0 \ll 1$, this region is of negligible importance.
Placing ourselves in a region where $\psi$ is small, {\it i.e.},
somewhere in the regions $r \lesssim r_0 - \sigma$ or
$r \gtrsim r_0 +\sigma$, we conclude that both marginally trapped 
surface equations
become
\begin{equation}
\label{eq:approximate trapped-surface equation}
f(r) \approx \left\{
\begin{array}{ll}
\displaystyle
1 + M_{\rm ADM} / 2r_0
& \mbox{\rm for $r \lesssim r_0 - \sigma$, while} \\
\displaystyle
1 - M_{\rm ADM} / 2r
& \mbox{\rm for $r \gtrsim r_0 + \sigma$.} \\
\end{array}
\right.
\end{equation}
We have neglected terms of order $\psi^2$.
We conclude from~(\ref{eq:approximate trapped-surface equation}) that if
the {\sc adm} mass of the gravitational system is sufficiently small,
that is, if $M_{\rm ADM} < 2r_0$, then $f(r)$ is strictly positive, and
no marginally trapped surface exists.
On the other hand, if $M_{\rm ADM} > 2r_0$, then $f(r)$ crosses the
axis at $r = M_{\rm ADM} / 2$, and there exists a marginally trapped
surface.
In~\cite{bib:Clayton Demopoulos and Legare 1997}, it was found that
this inequality could be satisfied by a large range of 
(physically admissible)
values of the
parameters $\{A, r_0, \sigma\}$.
However, note that this analysis is not valid for 
$M_{\rm ADM} \approx 2r_0$, for then the marginally trapped surface would
reside at $r\sim r_0$, while we have stated from the outset that our
approximations are not valid in such a {\it r\'egime}.

Recently, there has been some controversy on the formation of black holes
in {\sc ngt} 
(see~\cite{bib:Burko and Ori Phys Rev Lett 75 2455 1995,%
bib:Moffat and Sokolov 1995}).
Although our results will eventually help in resolving these issues,
we should point out that they do not comprise a proof that black holes
indeed form in a generic stellar collapse.
We expect that our current numerical studies will answer these questions.

\acknowledgments

We thank M.\ A.\ Clayton,
J.\ W.\ Moffat, and P.\ Savaria for useful discussions.
This work was partially
funded by the Natural Sciences and Engineering
Research Council of Canada.
J.\ L\'egar\'e would like to thank the Walter C.\ Sumner Foundation
for their funding of this research.

\end{document}